%% file: majorana_decoherence.tex
\documentclass[12pt]{iopart}
\input{settings}

\usepackage{iopams}
\usepackage{setstack}

\begin{document}
\title{Bath-induced decoherence in finite-size Majorana wires at non-zero temperature}

%%%%%%%%%%%%%%%%%%%%%%%%%%%%%%%%%%%%%%%%%%%%%%%%%%%%%%%%%%%%%%%%%%%%%%%%%%%%%%%%
%% 						AFFILIATIONS        					    		  %%
%%%%%%%%%%%%%%%%%%%%%%%%%%%%%%%%%%%%%%%%%%%%%%%%%%%%%%%%%%%%%%%%%%%%%%%%%%%%%%%%

\author{Niels Breckwoldt$^{1,2,3}$, Thore Posske$^{3,4}$ and Michael Thorwart$^{3,4}$}

\address{$^1$Center for Free-Electron Laser Science CFEL, Deutsches Elektronen-Synchrotron DESY, Notkestra{\ss}e 85, 22607 Hamburg, Germany}
\address{$^2$Department of Physics, Universit{\"a}t Hamburg, Notkestra{\ss}e 9-11, 20607 Hamburg, Germany}
\address{$^3$The Hamburg Centre for Ultrafast Imaging, Luruper Chaussee 149, 22761 Hamburg, Germany}
\address{$^4$I. Institut f{\"u}r Theoretische Physik, Universit{\"a}t Hamburg, Notkestra{\ss}e 9-11, 20607 Hamburg, Germany}

%%%%%%%%%%%%%%%%%%%%%%%%%%%%%%%%%%%%%%%%%%%%%%%%%%%%%%%%%%%%%%%%%%%%%%%%%%%%%%%%
%% 					    ABSTRACT             					    		  %%
%%%%%%%%%%%%%%%%%%%%%%%%%%%%%%%%%%%%%%%%%%%%%%%%%%%%%%%%%%%%%%%%%%%%%%%%%%%%%%%%

\begin{abstract}
	Braiding Majorana zero-modes around each other is a promising route towards topological quantum computing. Yet, two competing maxims emerge when implementing Majorana braiding in real systems: 
	On the one hand, perfect braiding should be conducted adiabatically slowly to avoid non-topological errors. On the other hand, braiding must be conducted fast such that decoherence effects introduced by the environment are negligible, which are generally unavoidable in finite-size systems. This competition results in an intermediate time scale for Majorana braiding that is optimal, but generally not error-free.
	Here, we calculate this intermediate time scale for a T-junction of short one-dimensional topological superconductors coupled to a bosonic bath that generates fluctuations in the local electric potential, which stem from, e.g., environmental photons or phonons of the substrate. We thereby obtain boundaries for the speed of Majorana braiding with a predetermined gate fidelity.
	Our results emphasize the general susceptibility of Majorana-based information storage in finite-size systems and can serve as a guide for determining the optimal braiding times in future experiments.
\end{abstract}

%%%%%%%%%%%%%%%%%%%%%%%%%%%%%%%%%%%%%%%%%%%%%%%%%%%%%%%%%%%%%%%%%%%%%%%%%%%%%%%%
%% 					    INTRODUCTION         					    		  %%
%%%%%%%%%%%%%%%%%%%%%%%%%%%%%%%%%%%%%%%%%%%%%%%%%%%%%%%%%%%%%%%%%%%%%%%%%%%%%%%%

\section{Introduction}
	\label{sec:introduction}

	Majorana zero-modes are half-fermionic non-Abelian anyons \cite{Kitaev2001}, which have supposedly been detected in semiconducting wires in proximity to superconductors \cite{Deng2012,Rokhinson2012,Mourik2012,Das2012}, in atomic chains of transition metals on elementary superconductors \cite{Schneider2021,Schneider2020a,Kim2018}, and in superconducting vortices on special systems, e.g., Fe-based superconductors \cite{Wang2018,Chiu2020,Yin2015,Machida2019,Zhu2019,Liu2018}.
	The discovery of Majorana signatures in these systems followed their theoretical prediction \cite{Kitaev2001,Sau2010,Alicea2010,Lutchyn2010,Oreg2010,Alicea2011}. Yet, the experimental evidence for the existence of Majorana zero-modes is not conclusive and remains experimentally challenging. Several setups for the detection of Majorana zero-modes have been proposed, including electronic correlation measurements \cite{Haim2015,Gorski2018,Zhang2019b,Gong2020},  statistical noise correlation measurements \cite{Manousakis2020}, or local transport experiments on quantum dots \cite{Prada2017}.
	However, ultimate evidence to confirm the detection of Majorana quasiparticles can only be delivered by braiding experiments or equivalent approaches \cite{Vijay2016}. The anyonic character of Majorana zero-modes then becomes manifest in the Berry phase difference $\frac{\pi}{2}$ between the states of different fermionic parity. With two additional auxiliary Majorana quasiparticles, the operation becomes a quantum gate, which in principle enables non-universal protected quantum computing \cite{Nayak2008}. 
	The practical aspects of the topological protection in this setup remain debated, though. Unarguably, the existence of the zero-modes is topologically protected, but this does not necessarily extend to the quantum gate operation if electronic excitations lie close to the Fermi energy or if the Majorana zero-modes are spatially not sufficiently separated \cite{Budich2012,Pedrocchi2015,Mishmash2020,Sekania2017}.
	\newline
	For the fidelity of Majorana braiding, two mechanisms are of major importance. First, a realistic exchange of two Majorana quasiparticles takes a finite time. This deviation from an adiabatic exchange generally excites unwanted quasiparticles and alters the braiding phase \cite{Karzig2013,Scheurer2013}, except for designed finite-time schemes for Majorana braiding \cite{Posske2020}. For perfectly smooth braiding protocols, the resulting nonadiabatic errors decay exponentially with increasing braiding time $t_{\mathrm{b}}$ \cite{Nag2019,Tutschku2020,Tutschku2016}. Hence, the braiding time should be as long as possible to minimize nonadiabatic errors.
	Second, each realistic system is coupled to a bath with a certain temperature, which introduces fluctuations in the local chemical potentials. To reduce resulting braiding errors -- which necessarily arise if the Majorana wave functions spatially overlap \cite{Knapp2018} -- the braiding time should be as short as possible, in contrast to above. Previous studies of Majorana zero-modes coupled to a bosonic bath suggest that bath-induced braiding errors follow an inverse power law with respect to the braiding time \cite{Nag2019,Zhang2019c,Knapp2016}. Yet, these studies neglect thermal excitations while focusing on dephasing and relaxation in a minimal model of four Majorana zero-modes.
	Decoherence of Majornana zero-modes in presence of a bath, in particular in the context of Majorana qubits, has been studied using a similar theoretical framework as the present work based on a quantum master equation \cite{Steiner2020,Munk2020,Munk2019,Qin2019}. These studies have demonstrated the general susceptibility of Majorana zero-modes to environment-induced potential fluctuations in finite systems.
	\par
	In this work, we concentrate on the intermediate time scale for optimal Majorana braiding in finite chains that is generated by the competition of the above mentioned effects.
	To demonstrate this, we first numerically simulate the finite-time braiding of Majorana zero-modes with a T-junction of short Kitaev chains \cite{Kitaev2001} whose Majorana zero-modes are controlled by local gate potentials \cite{Alicea2011,Tutschku2016,Tutschku2020,Sekania2017}, and find the minimal times where the probability for quasiparticle excitations $\Gamma$, also called quasiparticle poisoning, is below a given threshold. For this we consider several threshold values between $5\,\%$ and $0.1\,\%$, which we regard as realistic proof-of-principle values in forthcoming braiding experiments.
	Next, we couple a bosonic bath to the Majorana T-junction. By employing the Bloch-Redfield equation, we determine the maximal times for the same probability for quasiparticle poisoning. As a result, we obtain time frames where Majorana braiding is below a specified fraction of quasiparticle excitations in dependence on the temperature and the strength of the coupling to the bath.
	\newline
	The boundaries for the braiding time -- defined by the competition of nonadiabatic braiding errors and thermal quasiparticle excitations -- are an orientation for the next experimental milestone of braiding Majorana zero-modes for the first time.
	Additionally, this procedure is important for eventual technological applications of Majorana braiding for Majorana-based quantum computing, where $\Gamma$ should be well below $10^{-6}$, such that quantum error correction schemes can be applied.
	\par
	The structure of this manuscript is as follows:
	In \cref{sec:kitaev_model}, we introduce the model for the one-dimensional topological superconductors that are coupled to a bosonic bath.
	The resulting minimal time for Majorana braiding with a specified $\Gamma$ is determined in \cref{sec:braiding}, while the maximal time is determined in \cref{sec:dissipation}. Both results are combined to find the optimal time scales for Majorana braiding in \cref{sec:combined_error}. We conclude this work in \cref{sec:conclusion}.

%%%%%%%%%%%%%%%%%%%%%%%%%%%%%%%%%%%%%%%%%%%%%%%%%%%%%%%%%%%%%%%%%%%%%%%%%%%%%%%%
%% 					    MODEL                					    		  %%
%%%%%%%%%%%%%%%%%%%%%%%%%%%%%%%%%%%%%%%%%%%%%%%%%%%%%%%%%%%%%%%%%%%%%%%%%%%%%%%%

\section{Model}
\label{sec:kitaev_model}

\begin{figure}
	\centering
	\includegraphics[width=\textwidth]{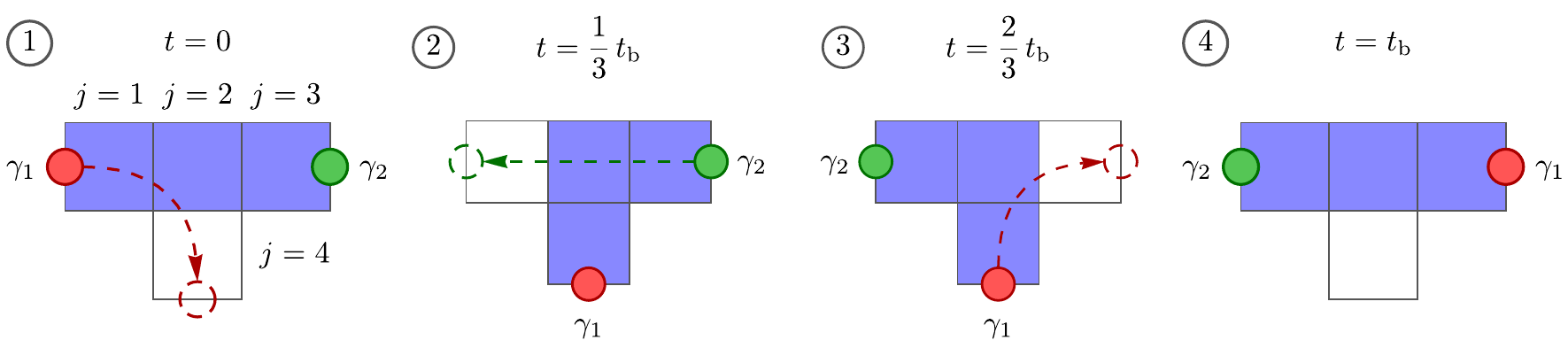}
	\caption{Braiding scheme for a 4-site Kitaev T-junction. Initially, the horizontally aligned sites are tuned into the topologically non-trivial parameter regime (blue) hosting Majorana quasiparticles $\gamma_{1}$, $\gamma_{2}$, whereas the vertical branch is in the topologically trivial regime (white). By individually switching the sites between both phases, the Majorana quasiparticles are moved on the T-junction, as indicated by the dashed arrows. Finally, the initial Hamiltonian is recovered, but both Majorana quasiparticles have been exchanged.}
	\label{fig:braiding_scheme}
\end{figure}

	Majorana zero-modes appear as zero-energy quasiparticle excitations at the ends of topological superconductors \cite{Leijnse2012,Alicea2012,Beenakker2013}. In second quantization, they are described by hermitian operators $\gamma_{j} = \gamma_{j}^{\dagger}$, fulfilling the anti-commutation relation, 
	\begin{equation}
	\label{eq:anti_commutation}
		\{ \gamma_{i}, \gamma_{j} \} = 2 \delta_{ij},
	\end{equation}
	similar to ordinary (Dirac) fermions. The Kitaev chain \cite{Kitaev2001} is a 1D toy model of a spinless $p$-wave superconductor that can host Majorana zero-modes. The Hamiltonian of an $N$-site Kitaev chain reads
	\begin{equation}
	\label{eq:kitaev_model}
		H_{\mathrm{K}} = - \sum_{j=1}^{N-1} \big( \epsilon c_j^{\dagger} c_{j+1} + \abs{\Delta} \e{\ii \phi} c_j c_{j+1} + h.c. \big) + \sum_{j=1}^{N} \mu_j c_j^{\dagger} c_j,
	\end{equation}
	where, $c_i^{(\dagger)}$ is a spinless fermion annihilation (creation) operator and $\mu_j$, $\epsilon$, and $\abs{\Delta} \e{\ii\phi}$ denote the local electric potential at site $j$, the electronic hopping strength and the superconducting pairing potential, respectively.
	\newline
	The Kitaev model exhibits two distinct topological phases: the topologically trivial phase and a topologically non-trivial one. In finite systems, the latter is characterized by an energy gap that separates the two lowest eigenstates from the residual states. The parameter choice $\mu_j = 0,\ \forall j$, and $ \abs{\Delta} = \epsilon \neq 0$ constitutes a special point in parameter space. At this point, \cref{eq:kitaev_model} can be written as
	\begin{equation}
	\label{eq:kitaev_topological_phase}
		H_{\mathrm{K}} = -\ii \epsilon \sum_{j=1}^{N-1} \gamma_{2j} \gamma_{2j+1},
	\end{equation}
	with Majorana operators
	\begin{eqnarray}
	\label{eq:majorana_1}
		\gamma_{2j} &= \e{-\ii \frac{\phi}{2}} c_{j}^{\dagger} + h.c., \\
	\label{eq:majorana_2}
		\gamma_{2j-1} &=  \ii \mathrm{e}^{-\ii \frac{\phi}{2}} c_{j}^{\dagger} + h.c. 
	\end{eqnarray}
	These operators fulfil the defining anti-commutation relation \cref{eq:anti_commutation}. In terms of fermion operators,
	\begin{equation}
		d_{i} = \frac12 \left( \gamma_{2j+1} + \ii \gamma_{2j} \right),
	\end{equation}
	\cref{eq:kitaev_topological_phase} reads
	\begin{equation}
	H_{\mathrm{K}} = \epsilon \sum_{j=1}^{N-1} \left( 2 d_{j}^{\dagger} d_{j} -1 \right). 
	\end{equation}
	The absent Majorana operators $\gamma_{1}$ and $\gamma_{2N}$ at both ends of the chain result in a non-local, zero-energy fermion, 
	\begin{equation}
	d_{\mathrm{end}} = \frac12 \left( \gamma_{1} + \ii \gamma_{2N} \right).
	\end{equation}
	This gives rise to a two-fold degenerate ground state, $\Ket{0}$ and $\Ket{1} = d_{\mathrm{end}}^{\dagger} \Ket{0}$ , where $d_{\mathrm{end}} \Ket{0} = 0$.
	\newline
	Majorana quasiparticles can be moved by individually tuning each site between the topologically trivial and non-trivial regime, for instance, by a variable electric field \cite{Alicea2011,Tutschku2020,Sekania2017}. For exchanging Majorana zero-modes, we investigate networks of 1D wires, although braiding without the usage of such networks is in general possible \cite{Alicea2011,Landau2016,Plugge2017,Bonderson2009,Bonderson2008,Vijay2016,Chiu2015}. We employ a modified Kitaev model for a $4$-site T-junction as the simplest possible realisation of a wire network. Throughout this work, we choose $\abs{\Delta} =~\epsilon$.  As discussed in detail in \cite{Alicea2011}, the superconducting phases have to be chosen relative to a particular direction. Especially at the branching point of the T-junction, the phase difference must not equal an integer multiple of $\pi$ to avoid the wire segments to decouple \cite{Alicea2011}. With $\varphi = 0$ and $\varphi = \pi / 2$ in the horizontal and vertical wire, respectively, the Hamiltonian of the T-junction reads
\begin{equation}
	\label{eq:hamiltonian_T-junction}
	\fl
		H_{\mathrm{T}}(t)= - \epsilon \sum_{j=1}^{2} \left(c_j^{\dagger} c_{j+1} + c_j c_{j+1} + h.c. \right) - \epsilon \left(c_{4}^{\dagger} c_{2} + \ii c_{4} c_{2} + h.c. \right) + \sum_{j=1}^{4} \mu_{j}(t) c_j^{\dagger} c_j.
\end{equation}
 	Here, we introduced time-dependent local electric potentials $\mu_{j}(t)$ that are specified in \cref{sec:braiding}.\\	
	Since the Hamiltonian in \cref{eq:hamiltonian_T-junction} conserves fermionic parity, it is convenient to choose pure parity states, $\ket{\pm}$, as equivalent to the degenerate ground state $\ket{0}$, $\ket{1}$, where the sign $\pm$ refers to the ground state of even and odd parity, respectively.

%%%%%%%%%%%%%%%%%%%%%%%%%%%%%%%%%%%%%%%%%%%%%%%%%%%%%%%%%%%%%%%%%%%%%%%%%%%%%%%%
%% 					    BRAIDING-INDUCED ERROR 					    		  %%
%%%%%%%%%%%%%%%%%%%%%%%%%%%%%%%%%%%%%%%%%%%%%%%%%%%%%%%%%%%%%%%%%%%%%%%%%%%%%%%%

\section{Braiding-induced error}
\label{sec:braiding}

	\begin{figure}
		\centering
		\includegraphics[width=\textwidth]{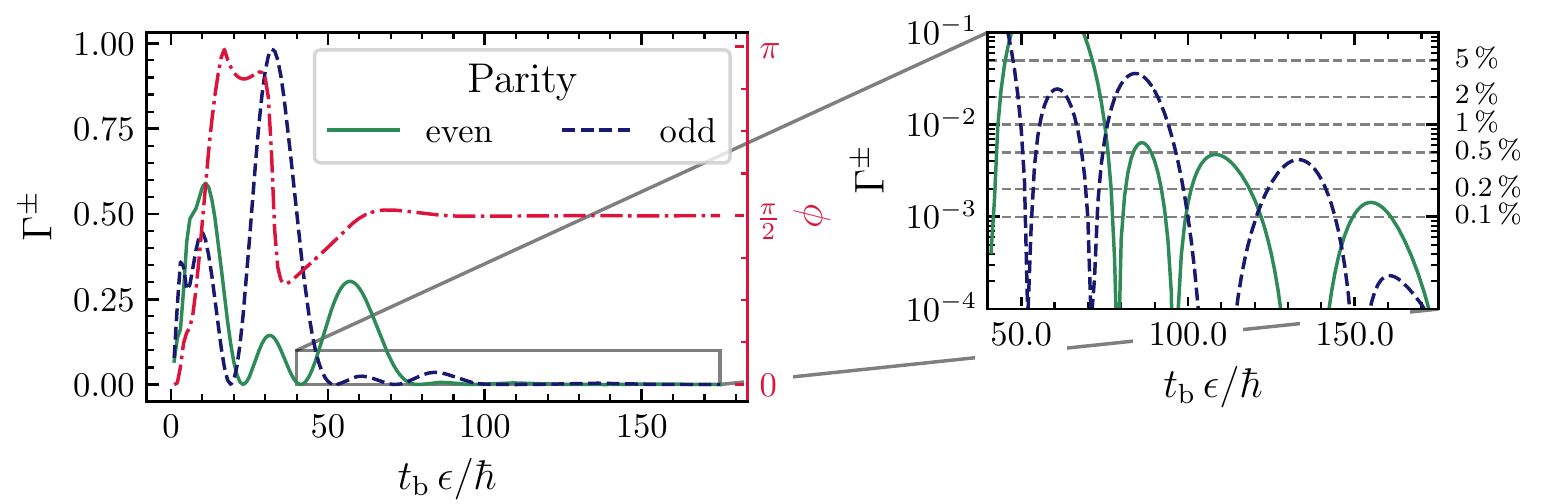}
		\caption{Quasi-particle excitations $\Gamma^{\pm}$ in dependence on the finite braiding time $t_{\mathrm{b}}$. The dashed-dotted red curve refers to the relative phase difference $\phi$ that converges to the adiabatic value of $\pi/2$.}
		\label{fig:braiding_decoherence}
	\end{figure}

	For braiding Majorana quasiparticles, we consider an exchange protocol with three distinct steps \cite{Alicea2011,Tutschku2020}, as illustrated in \cref{fig:braiding_scheme}. To that purpose, we specify the local electric potentials as
	\begin{equation}
		\mu_{j}(t) = \mu_{j}^{(0)} + \delta\mu_{j}(t),
	\end{equation}
	with $\delta\mu_{j}(t)$ denoting deviations from the initial electric potentials $\mu_{1}^{(0)} = \mu_{2}^{(0)} = \mu_{3}^{(0)} = 0$ and $\mu_{4}^{(0)} = \epsilon$. For $\delta\mu_{j}(t)$, we consider smooth functions, similar to those in \cite{Wieckowski2020},
	\begin{eqnarray}
		\delta\mu_1(t) = \epsilon \cdot \sin^2\left[ \frac{\pi}{2} \tri \left( 3 \frac{t}{t_\mathrm{b}} - 1 \right) \right],\\
		\delta\mu_2(t) = 0,\\
		\delta\mu_3(t) = \epsilon \cdot \sin^2 \left[ \frac{\pi}{2} \tri \left(3 \frac{t}{t_\mathrm{b}} - 2\right) \right],\\
		\delta\mu_4(t) = \epsilon \cdot \cos^2 \left[ \frac{\pi}{2} \left( \tri (3 \frac{t}{t_\mathrm{b}} - 1) + \tri (3 \frac{t}{t_\mathrm{b}} - 2) \right) \right]
	\end{eqnarray}
	with the triangle function
	\begin{equation}
		\tri(x) \overset{\mathrm{def}}{=} \max( 1- \abs{x}, 0).
	\end{equation}
	The braiding time, i.e., the time for an exchange of Majorana zero-modes, is denoted by $t_{\mathrm{b}}$.
	Note that individual sites are locally tuned into the topologically trivial regime by changing their local electric potential to $\mu_{j} = \epsilon$ while maintaining finite electronic hopping and the superconducting pairing potential, as done previously \cite{Alicea2011,Sekania2017,Tutschku2020}. The system as a whole, however, remains topologically non-trivially during the exchange as the energy gap does not close.\\
	In order to determine the minimal necessary time for a low-error exchange of Majorana zero-modes, we regard the fraction of quasi-particle excitations caused by the finite braiding time $t_{\mathrm{b}}$,
	\begin{equation}
		\Gamma^{\pm}(t_{\mathrm{b}}) = 1 - \abs{\tau^{\pm}(t=t_\mathrm{b})}^{2},
	\end{equation}
	with the transition amplitude $\tau^{\pm}(t) = \bra{\pm} U(t) \ket\pm$. We consider the exemplary fractions of quasi-particle excitations of $0.1\,\%$, $0.2\,\%$, $0.5\,\%$, $1\,\%$, $2\,\%$ and $5\,\%$ as reasonable boundaries for the first experimental realization of Majorana braiding. For technical applications, however, the fidelity must fall below these numbers by orders of magnitude. The time evolution operator $U(t)$ is determined by numerically solving the Schr{\"o}dinger equation
	\begin{equation}
		\frac{\mathrm{d}}{\mathrm{d}t} U(t) = - \frac{\ii}{\hbar} H_{\mathrm{T}}(t) U(t), 
	\end{equation}
	with $\ U(0) = \mathbb{1}$.
	\newline
	As shown in \cref{fig:braiding_decoherence}, the fidelity-dependent braiding times strongly depend on the parity of the states and lie between $160.8\,\hbar / \epsilon$  for a quasi-particle excitation of $0.1\,\%$ and $47.6\,\hbar / \epsilon$ for $\Gamma = 5\,\%$, see \refTab{tab:min_braiding_time} for details. Additionally, the relative phase difference 
	\begin{equation}
		\phi(t_{\mathrm{b}}) = \abs{\arg \left( \tau^{+}(t_\mathrm{b}) \right) - \arg \left( \tau^{-}(t_\mathrm{b}) \right)}
	\end{equation}
	is shown in \cref{fig:braiding_decoherence} (see ordinate on the right-hand side). Since both parity sub-spaces remain degenerate during the Majorana exchange, $\phi$ reflects the difference of dynamic phases which converge towards the Berry phases $\pi/2$ in the adiabatic limit, revealing their non-Abelian nature.
	
	\begin{table}[h!]
		\centering
		\renewcommand{\arraystretch}{1.5}
		\newcommand{\tab}{\hspace{-5pt}}
		\begin{tabular}{@{}ccccccc@{}}
			\toprule
			$\Gamma\ (\%)$ \tab & $0.1$ \tab & $0.2$ \tab & $0.5$ \tab & $1$ \tab & $2$ \tab & $5$ \tab\\ 
			\midrule[0.75pt]
			$t_{\mathrm{b}}^{(+)}\ (\hbar / \epsilon)$ \tab & $160.8$ \tab & $118.5$ \tab & $89.0$ \tab & $75.1$ \tab & $73.8$ \tab & $71.4$ \tab\\
			$t_{\mathrm{b}}^{(-)}\ (\hbar / \epsilon)$ \tab & $144.1$ \tab & $141.1$ \tab & $96.4$ \tab & $94.1$ \tab & $91.0$ \tab & $47.6$ \tab\\
			$\vartheta_{\Gamma}\ (\epsilon / k_\mathrm{B})$ \tab & $0.099$ \tab & $0.110$ \tab & $0.129$ \tab & $0.149$ \tab & $0.176$ \tab & $0.232$ \tab\\
			\bottomrule	
		\end{tabular}
		\caption{Minimal exchange times $t_{\mathrm{b}}^{(\pm)}$ for a given fraction of quasi-particle excitations $\Gamma$ (also see \cref{fig:braiding_decoherence}). The sign $\pm$ denotes the sectors of even and odd fermionic parity, respectively. Also listed are the temperatures $\vartheta_{\Gamma}$ that correspond to the respective quasi-particle excitations of the Boltzmann distribution in thermal equilibrium (also cf. see \cref{sec:dissipation}).}
		\label{tab:min_braiding_time}
	\end{table}

%%%%%%%%%%%%%%%%%%%%%%%%%%%%%%%%%%%%%%%%%%%%%%%%%%%%%%%%%%%%%%%%%%%%%%%%%%%%%%%%
%% 					    BATH-INDUCED ERROR   					    		  %%
%%%%%%%%%%%%%%%%%%%%%%%%%%%%%%%%%%%%%%%%%%%%%%%%%%%%%%%%%%%%%%%%%%%%%%%%%%%%%%%%

\section{Bath-induced error}
\label{sec:dissipation}	

	\begin{figure}
		\centering
		\includegraphics[width=\textwidth]{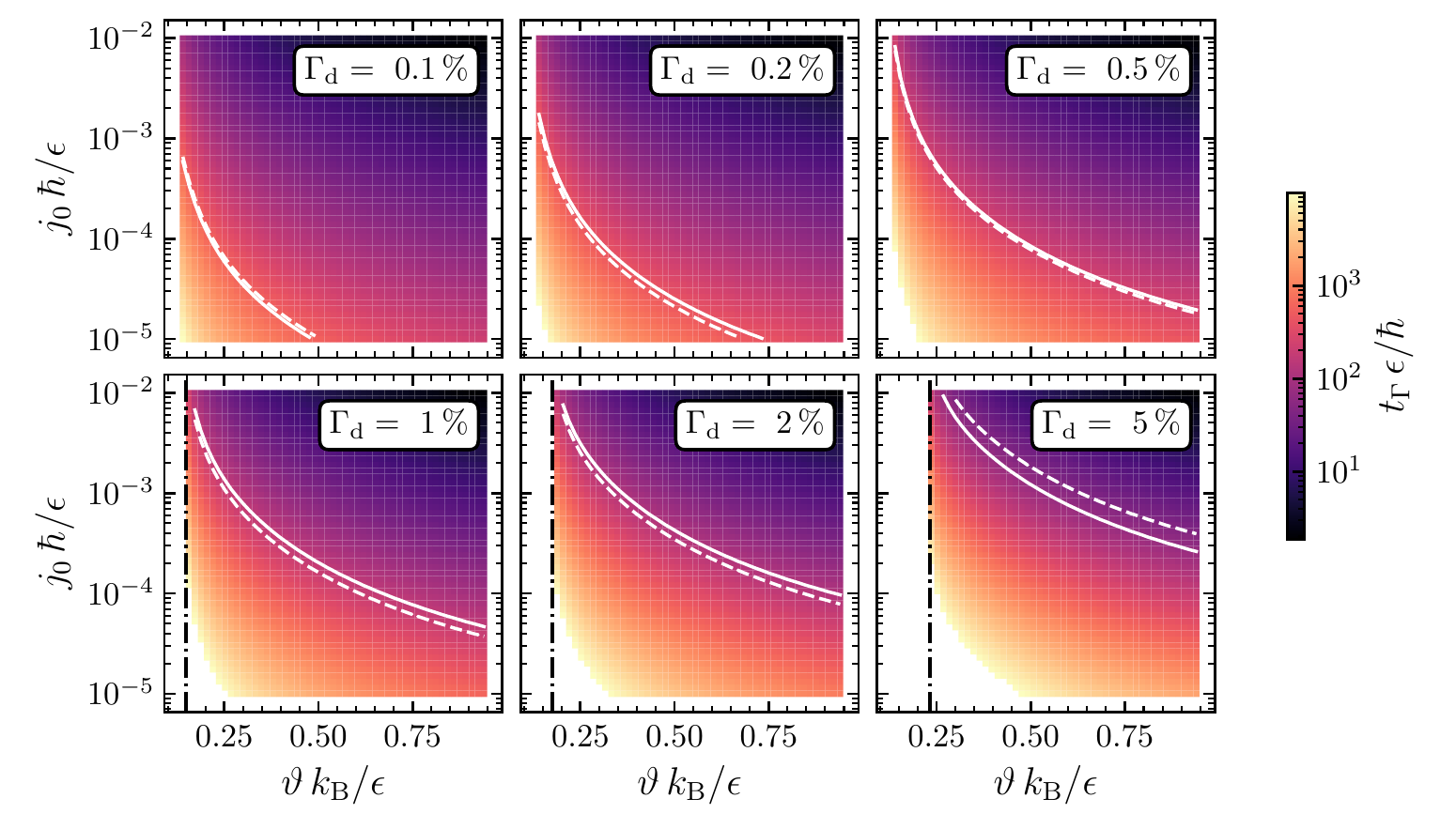}
		\caption{Dissipation time $t_{\Gamma}$ in dependence on temperature $\vartheta$ and system-bath coupling $j_{0}$ for given quasi-particle excitations $\Gamma_{\mathrm{d}}$ between $0.1\,\%$ and $5\,\%$. The blank areas indicate either a dissipation time that exceeds the simulated time span of $t_{\Gamma} > 10^{4}\,\hbar / \epsilon$ or a temperature below the threshold temperature $\vartheta_{\Gamma}$ where $t_{\Gamma} \rightarrow \infty$ (dash-dotted line; see text). Also shown are the contour lines from \refTab{tab:min_braiding_time} for even (solid) and odd (dashed) parity (also cf. \cref{fig:param_space}).}
		\label{fig:dissipation}
	\end{figure}

	For the description of environment-induced decoherence by fluctuating local electric potentials, we employ a system-bath model \cite{Breuer2007,Weiss1999,May2003} with a general Hamiltonian of the form
	\begin{equation}
		H = H_\mathrm{S} + H_\mathrm{B} + H_\mathrm{I},
	\end{equation}
	where $H_\mathrm{S,B}$, respectively, denote the Hamiltonian of the system and the bath, and $H_\mathrm{I}$ describes the interaction between both. The respective density matrices are denoted by $\rho_{\mathrm{S,B}}$.\\
	The system's Hamiltonian is chosen as the initial configuration of the T-junction in \cref{eq:hamiltonian_T-junction}, i.e. $ H_\mathrm{S} = H_{\mathrm{T}}(0)$. In our model, each lattice site is coupled to a non-interacting bath of harmonic oscillators:
	\begin{equation}
	H_{\mathrm{I}} = \sum_{j=1}^{4} c_{j}^{\dagger} c_{j} \sum_{k} \lambda_{k} \left( b_{k,j} + b_{k,j}^{\dagger} \right),
	\end{equation}
	with $b_{k,j}^{(\dagger)}$ and $\lambda_{k}$ denoting bosonic annihilation (creation) operators and the coupling constants, respectively. The system-bath interaction leads to fluctuations of the local electric potentials $\mu_{j}$. Such ubiquitous fluctuations can, e.g., stem from the varying electric field of thermal radiation or variations in the crystal fields by phonons. We neglect the possible influence of the bath on the tunneling amplitude, or superconductive pairing potential. The system-bath coupling is assumed to be weak, such that the bath stays in thermal equilibrium, i.e., $\frac{\partial}{\partial t} \rho_{\mathrm{B}} = 0$, known as Born approximation. By further assuming Markovian behaviour, i.e., $\rho_{\mathrm{S}}(t-t^{\prime}) \simeq \rho_{\mathrm{S}}(t)$, the time evolution of the system is governed by the Born-Markov quantum master equation \cite{Breuer2007,May2003,CohenTannoudji1998}
	\begin{equation}\label{eq:born_markov_master}
		\dot{\tilde{\rho}}_{\mathrm{S}}(t) =  \frac{-1}{\hbar^{2}}\int_{0}^{\infty} \dd{t'}  \tr{B} [ \tilde{H}_\mathrm{I}(t),
		[ \tilde{H}_\mathrm{I}(t - t'), \tilde{\rho}_{\mathrm{S}}(t) \otimes \rho_\mathrm{B}] ].
	\end{equation}
	Here, $\tr{\mathrm{B}}$ denotes the partial trace over the bath degrees freedom, and the tilde $(\tilde{\rho}_{\mathrm{S}},\ \tilde{H}_\mathrm{I})$ indicates operators in the interaction picture with respect to $H_{0} = H_{\mathrm{S}} + H_{\mathrm{B}}$.\\
	The effects of the bath are fully described by the bath correlation function $C(\omega)$, which in the frequency domain reads
	\begin{equation}
		C(\omega) = 2 \pi \left[ 1 + n(\omega) \right] \left[ J(\omega) - J(-\omega) \right],
	\end{equation}
	with $J(\omega)$ being the spectral density and $n(\omega)$ the Bose-Einstein distribution at given temperature $\vartheta$,
	\begin{equation}
		n(\omega) = \left[ \exp \left( \frac{\hbar \omega}{k_{\mathrm{B}} \vartheta} \right) - 1 \right]^{-1}.
	\end{equation}
	Here, $k_{\mathrm{B}}$ denotes the Boltzmann constant. For a sufficiently large number of bath modes, we assume the spectral density to be continuous \cite{May2003}. Here, we employ a spectral density of Ohmic type with an exponential cut-off,
	\begin{equation}
		J(\omega) = \Theta(\omega) j_{0} \omega \e{-\omega / \omega_{\mathrm{c}}},
	\end{equation}
	where the constant $j_{0}$ characterizes the coupling strength, and $\Theta(\omega)$ denotes the Heaviside step function. In all our simulations, the cut-off frequency is $\omega_{\mathrm{c}} = 100\,\epsilon / \hbar$.\\
	Evaluating \cref{eq:born_markov_master} in the basis of eigenstates of $H_\mathrm{S}$, denoted by Latin subscripts, yields the Bloch-Redfield equation \cite{May2003,CohenTannoudji1998}
	\begin{equation}
	\label{eq:redfield}
		\left(\dot{\rho}_{\mathrm{S}} \right)_{ab}(t) = - \ii \omega_{ab} \left(\rho_{\mathrm{S}} \right)_{ab}(t) - \sum_{c,d} R_{ab,cd} \left(\rho_{\mathrm{S}} \right)_{cd}(t).
	\end{equation}
	Here, $ R_{ab,cd}$ is the Redfield tensor
	\begin{equation}
		\fl
		 R_{ab,cd} =\ \delta_{ac} \sum_{e} \Lambda_{be,ed}(\omega_{de}) + \delta_{bd} \sum_{e} \Lambda_{ae,ec}(\omega_{ce}) - \Lambda_{db,ac}(\omega_{ca}) - \Lambda_{ca,bd}(\omega_{db}),
	\end{equation}
	with the damping tensor
	\begin{equation}
		\Lambda_{ab,cd}(\omega) = \Real \left[ \int_{0}^{\infty} \dd{t^{\prime}} C(t^{\prime}) \e{\ii \omega t^{\prime}} \sum_{i,j = 1}^{4} (c_{i}^{\dagger} c_{i})_{ab} (c_{j}^{\dagger} c_{j})_{cd} \right].
	\end{equation}
	As in \cref{sec:braiding}, we consider six different magnitudes of quasiparticle excitations $\Gamma$, but now vary the temperature and strength of the coupling to the bath. In the considered Born-Markov framework, dissipation only depends on energy differences rather than on the corresponding states. Therefore, we focus on the even parity subspace only, as parity conservation is not broken by the bath. We numerically solve \cref{eq:redfield} with a matrix exponential approach and the initial condition $\rho_{\mathrm{S}}(0) = \ket{+} \bra{+}$. We further define quasi-particle excitations induced by dissipation as
	\begin{equation}
	\Gamma_{\mathrm{d}}(t) = 1 - \braket{ + \vert \rho_{\mathrm{S}}(t) \vert +}.
	\end{equation}
	The results are shown in \cref{fig:dissipation}, where $t_{\Gamma}$ denotes the time until the given fraction of quasi-particle excitation is reached. In thermal equilibrium, the occupation probability $p_i$ of the respective microstate $i$ necessarily approximates the Boltzmann distribution
	\begin{equation}
		\label{eq:boltzmann}
		p_i = \frac{1}{Z} \exp\left( - \frac{E_i}{k_\mathrm{B} \vartheta} \right),
	\end{equation}
	where $E_i$ are the many-particle eigenenergies of the Hamiltonian in \cref{eq:hamiltonian_T-junction} and $Z$ denotes the canonical partition function
	\begin{equation}
		Z(\vartheta) = \sum_{i} \exp\left(- \frac{E_i}{k_\mathrm{B} \vartheta} \right).
	\end{equation}
	Here, the sum runs over all possible microstates.
	Evaluated for the ground state of the system, \cref{eq:boltzmann} defines a unique mapping  between a given amount of quasi-particle excitations $\Gamma$ in thermal equilibrium and an associated temperature $\vartheta_\Gamma$, which we denote as threshold temperature,
	\begin{equation}\label{eq:margin_temp}
		Z(\vartheta_\Gamma) \left( 1 - \Gamma \right) = 1.
	\end{equation}
	For temperatures below the threshold, thermal excitations will eventually result in less decoherence in thermal equilibrium, whereas temperatures above the threshold lead to more decoherence. The numerical values of $\vartheta_\Gamma$ are summarized in \refTab{tab:min_braiding_time}. These threshold temperatures are in good agreement with the Bloch-Redfield dissipation simulations at large times, i.e., in the thermodynamic limit in \cref{fig:dissipation} (dash-dotted line; note that the threshold temperatures for $0.1\,\%$, $0.2\,\%$ and $0.5\,\%$ are below the simulated temperature range and hence not drawn.).
	This serves as an important consistency check of the more general Bloch-Redfield results.
	In the low-temperature regime, i.e., $\vartheta \ll E_\mathrm{gap} / k_\mathrm{B}$, where $E_\mathrm{gap}$ denotes the energy gap between the degenerate ground state and the first excited state, it is sufficient to evaluate \cref{eq:margin_temp} by neglecting all energy levels above the energy gap, resulting in no significant changes of the threshold temperatures.

%%%%%%%%%%%%%%%%%%%%%%%%%%%%%%%%%%%%%%%%%%%%%%%%%%%%%%%%%%%%%%%%%%%%%%%%%%%%%%%%
%% 					    OPTIMAL BRAIDING TIME SCALE 			    		  %%
%%%%%%%%%%%%%%%%%%%%%%%%%%%%%%%%%%%%%%%%%%%%%%%%%%%%%%%%%%%%%%%%%%%%%%%%%%%%%%%%

\section{Optimal time scales for Majorana braiding}
\label{sec:combined_error}

	\begin{figure}
		\centering
		\includegraphics[width=\textwidth]{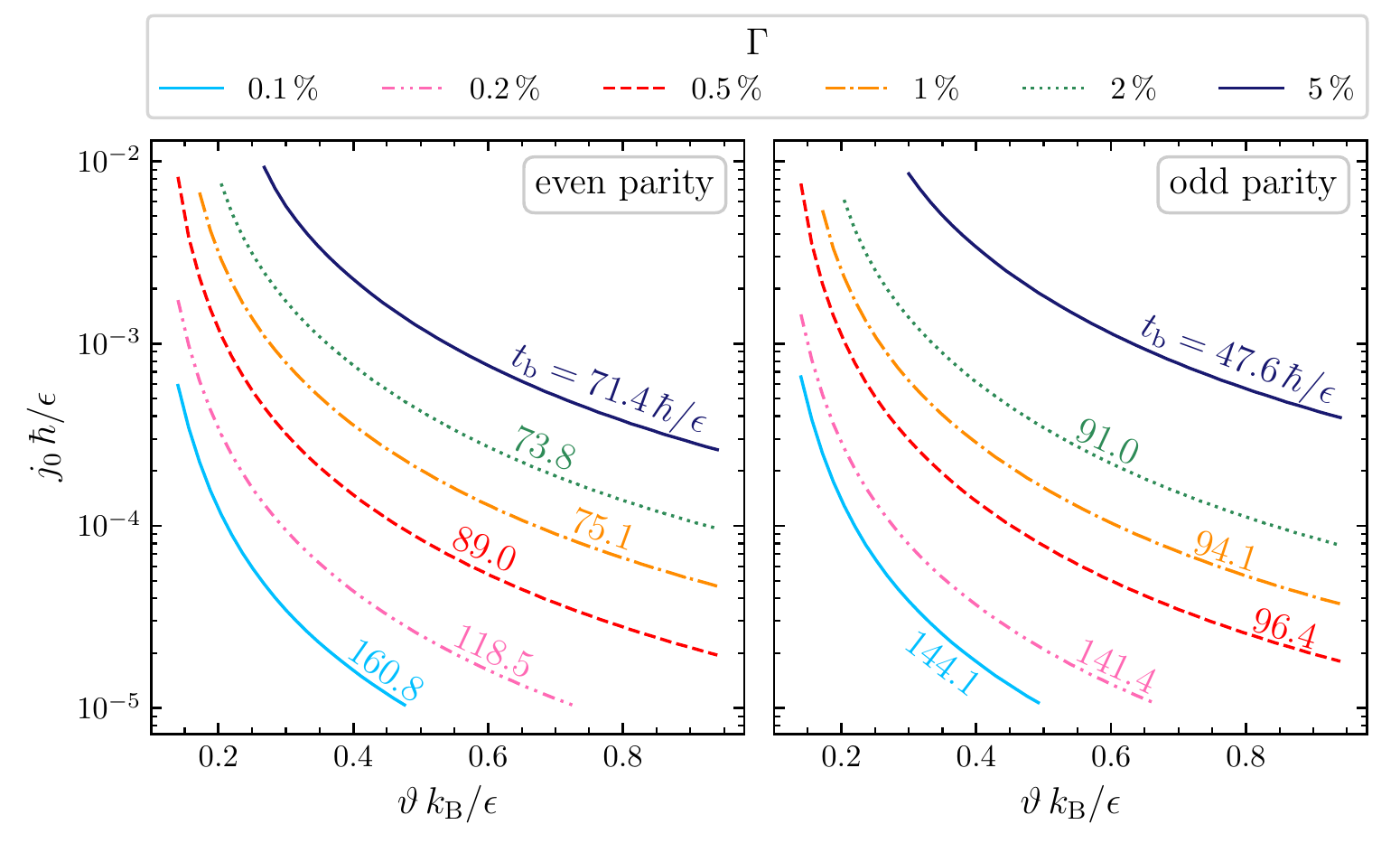}
		\caption{Curves of equal braiding- and dissipation time, $t_{\mathrm{b}}^{(\pm)} = t_{\Gamma}$, in the parameter space of temperature $\vartheta$ and system-bath coupling $j_{0}$, respectively, for the considered fraction of quasi-particle excitations $\Gamma$ between $0.1\,\%$ and $5\,\%$. The corresponding minimal braiding times $t_{\mathrm{b}}$ for even (top panel) and odd parity (bottom panel) are taken from \refTab{tab:min_braiding_time}.}
		\label{fig:param_space}
	\end{figure}

	In order to minimize the resulting braiding error, two competing principles emerge: On the one hand, perfect braiding needs to be adiabatically slow. On the other hand, the braiding time must be as short as possible to inhibit environment-induced dissipation. This competition results in an intermediate time scale where Majorana braiding is optimal, though not error-free.\\
	We can identify two regimes for each fraction of quasiparticle excitations in \cref{fig:dissipation} that are separated by contour lines $t_{\mathrm{b}}^{(\pm)} = t_{\Gamma}$, where, for a given fraction of quasiparticle excitations $\Gamma$, the dissipation time coincides with the fastest braiding time (see \refTab{tab:min_braiding_time}). These isolines, as depicted in \cref{fig:param_space}, respectively define the largest temperature and strongest system-bath interaction that result in an error smaller or equal to the targeted accuracy and constitute the fastest, hence optimal, time for braiding. Experimental conditions for which the dissipation time $t_{\Gamma}$ exceeds the fastest exchange time, i.e., the area in the parameter space of temperatures and couplings  below the corresponding curve, will reduce dissipation-induced errors, and define favourable conditions for braiding experiments to be performed at. The area above the curve, in contrast, represents unfavourable conditions as the dissipation-induced error exceeds the braiding-induced one and limits the fidelity.\\
	To apply this method to realistic Majorana systems, the low-energy model of the realised Majorana chain in combination with the coupling of the bath needs to be determined for two different temperatures. This low-energy model can be derived by a fit to the spatial spread and the energy gap of the Majorana zero-modes, while the coupling to the bath can be determined by measuring either the time for thermal equilibration or the rate with which excitations from the ground state are generated.
	In our model, the superconducting gap $E_{\mathrm{gap}}$ solely depends on the tunneling strength $\epsilon$, showing a constant ratio of $E_{\mathrm{gap}} / \epsilon \approx 0.7$, with the order of magnitude of the superconducting band gap typically being $E_{\mathrm{gap}}~\sim~10^{-4}\,$eV \cite{Mourik2012,Schneider2020a,Kim2018}. Here, we consider the superconducting gap as the relevant parameter as it determines the system's susceptibility to environment-induced dissipation by setting the protective energy scale. 
	This implies the fastest time for braiding to range between  $t_{\mathrm{b}} \approx 0.7\,$ns for an amount of quasi-particle excitations of $0.1\,\%$, and $0.2\,$ns for $\Gamma = 5\,\%$.
	The investigated temperature domain corresponds to temperatures between $\vartheta \sim 0.2\,E_{\mathrm{gap}} / k_{\mathrm{B}} \approx 0.2\,$K and $\sim 1.4\,E_{\mathrm{gap}} / k_{\mathrm{B}}~\approx~1.6\,$K.
	The system-bath coupling is on the order of $j_{0} \sim 15 \cdot 10^{-6}\, E_{\mathrm{gap}} / \hbar \approx 2 \cdot 10^{6}\,\mathrm{s}^{-1}$ and $\sim 15 \cdot 10^{-3}\, E_{\mathrm{gap}} / \hbar \approx 2 \cdot 10^{9}\,\mathrm{s}^{-1}$.

%%%%%%%%%%%%%%%%%%%%%%%%%%%%%%%%%%%%%%%%%%%%%%%%%%%%%%%%%%%%%%%%%%%%%%%%%%%%%%%
% 					    CONCLUSION          					    		  %%
%%%%%%%%%%%%%%%%%%%%%%%%%%%%%%%%%%%%%%%%%%%%%%%%%%%%%%%%%%%%%%%%%%%%%%%%%%%%%%%

\section{Conclusion}
\label{sec:conclusion}

	After the detection of strong signatures of Majorana zero-modes, the next milestone in Majorana physics is to braid Majorana quasiparticles around each other and to verify the characteristic change of the ground state. 
	To this end, it is important to maximize the braiding time in order to reach adiabatic conditions. On the other hand, minimizing the braiding time is necessary in order to reduce errors introduced by the environment. This competition leads to an intermediate time scale for each specific system that is optimal for Majorana braiding.
	We present a method and model calculations how to determine this intermediate time regime based on exact diagonalization of short T-junctions of 1D topological superconductors and the Bloch-Redfield equation. Our work shows that Majorana zero-modes in finite-size wires and chains suffer from thermal quasiparticle excitations when coupled to a bosonic bath. 
	The observed decoherence can in principle be a combined effect of two contributions. First, the potential fluctuations of the bath may effectively delocalize the Majorana zero-modes, leading to a finite Majorana overlap in the centre of the chain that facilitates decoherence by the bosonic environment. Although this effect is exponentially suppressed in longer chains, it remains finite in every finite-size system. Second, our work leaves open the possibility for a genuine decoherence of topologically localized Majorana modes due to the bosonic environment that challenges the point of view that Majorana zero modes are immune to bosonic decoherence even if their net overlap vanishes. To which extent each contribution is present requires further length-dependent studies of Majorana wires coupled to a bosonic bath that exceed the scope of this work demanding for further studies, in particular on the expected exponential decay of the first-mentioned contribution depending on an induced Majorana overlap.
	In our finite-size system, the characteristic time scale for dissipation coincides with the fastest braiding time that is necessary to achieve a given fidelity goal. This competition implies an optimal braiding time that strongly depends on experimental conditions, like, for instance, the temperature or coupling strength to the environment. Our results emphasize the susceptibility of Majorana-based information storage in finite-size systems, and the important role of dissipative effects in forthcoming braiding experiments.

\begin{ack}
	TP thanks Ching-Kai Chiu and Christian Tutschku for discussions. This work is supported by the Cluster of Excellence "CUI: Advanced Imaging of Matter" of the Deutsche Forschungsgemeinschaft (DFG) -- EXC 2056 -- project ID 390715994.
\end{ack}

% Bibliography
\section*{References}
\bibliographystyle{iopart-num}
\bibliography{majorana_decoherence}

\end{document}

%% file: settings.tex
\usepackage[utf8]{inputenc}
\usepackage[T1]{fontenc}

\usepackage{xcolor}

\usepackage{bbold}
\usepackage{braket}
\usepackage{graphicx}
\usepackage{pgf}
\usepackage{import}

\usepackage{hyperref}
\usepackage[capitalise]{cleveref}
\crefname{section}{Sec.}{}
\crefname{appendix}{App.}{}
\crefname{chapter}{Chap.}{Chaps.}

\newcommand{\tri}{\mathrm{tri}}
\renewcommand{\tr}[1]{\mathrm{tr}_\mathrm{#1}}
\newcommand{\ii}{\mathrm{i}}
\renewcommand{\e}[1]{\mathrm{e}^{#1}}
\newcommand{\dd}[1]{\hspace{-5pt} \mathrm{d}{#1\ }}
\newcommand{\Real}{\mathrm{Re}}
\newcommand{\abs}[1]{\vert #1 \vert}

\usepackage{booktabs}

\usepackage{color}

\newcommand{\refTab}[1]{Tab.~\ref{#1}}

%% file: majorana_decoherence.bbl
\providecommand{\newblock}{}
\begin{thebibliography}{10}
\expandafter\ifx\csname url\endcsname\relax
  \def\url#1{{\tt #1}}\fi
\expandafter\ifx\csname urlprefix\endcsname\relax\def\urlprefix{URL }\fi
\providecommand{\eprint}[2][]{\url{#2}}

\bibitem{Kitaev2001}
Kitaev A~Y 2001 {\em Phys. Usp.\/} {\bf 44} 131

\bibitem{Deng2012}
Deng M~T, Yu C~L, Huang G~Y, Larsson M, Caroff P and Xu H~Q 2012 {\em Nano
  Lett.\/} {\bf 12} 6414--6419

\bibitem{Rokhinson2012}
Rokhinson L~P, Liu X and Furdyna J~K 2012 {\em Nat. Phys.\/} {\bf 8} 795--799
  ISSN 1745-2481

\bibitem{Mourik2012}
Mourik V, Zuo K, Frolov S~M, Plissard S~R, Bakkers E~P~A~M and Kouwenhoven L~P
  2012 {\em Science\/} {\bf 336} 1003--1007

\bibitem{Das2012}
Das A, Ronen Y, Most Y, Oreg Y, Heiblum M and Shtrikman H 2012 {\em Nat.
  Phys.\/} {\bf 8} 887--895

\bibitem{Schneider2021}
Schneider L, Beck P, Neuhaus-Steinmetz J, Posske T, Wiebe J and Wiesendanger R
  2021 {\em arXiv:2104.11503\/} {\bf [cond-mat.supr-con]}

\bibitem{Schneider2020a}
Schneider L, Brinker S, Steinbrecher M, Hermenau J, Posske T, dos Santos~Dias
  M, Lounis S, Wiesendanger R and Wiebe J 2020 {\em Nat. Commun.\/} {\bf 11}
  4707 ISSN 2041-1723

\bibitem{Kim2018}
Kim H, Palacio-Morales A, Posske T, R{\'{o}}zsa L, Palot{\'{a}}s K, Szunyogh L,
  Thorwart M and Wiesendanger R 2018 {\em Sci. Adv.\/} {\bf 4} eaar5251

\bibitem{Wang2018}
Wang D, Kong L, Fan P, Chen H, Zhu S, Liu W, Cao L, Sun Y, Du S, Schneeloch J,
  Zhong R, Gu G, Fu L, Ding H and Gao H~J 2018 {\em Science\/} {\bf 362}
  333--335

\bibitem{Chiu2020}
Chiu C~K, Machida T, Huang Y, Hanaguri T and Zhang F~C 2020 {\em Sci. Adv.\/}
  {\bf 6} eaay0443

\bibitem{Yin2015}
Yin J~X, Wu Z, Wang J~H, Ye Z~Y, Gong J, Hou X~Y, Shan L, Li A, Liang X~J, Wu
  X~X, Li J, Ting C~S, Wang Z~Q, Hu J~P, Hor P~H, Ding H and Pan S~H 2015 {\em
  Nat. Phys.\/} {\bf 11} 543--546

\bibitem{Machida2019}
Machida T, Sun Y, Pyon S, Takeda S, Kohsaka Y, Hanaguri T, Sasagawa T and
  Tamegai T 2019 {\em Nat. Mater.\/} {\bf 18} 811--815

\bibitem{Zhu2019}
Zhu S, Kong L, Cao L, Chen H, Papaj M, Du S, Xing Y, Liu W, Wang D, Shen C,
  Yang F, Schneeloch J, Zhong R, Gu G, Fu L, Zhang Y~Y, Ding H and Gao H~J 2019
  {\em Science\/} {\bf 367} 189--192

\bibitem{Liu2018}
Liu Q, Chen C, Zhang T, Peng R, Yan Y~J, Wen C~H~P, Lou X, Huang Y~L, Tian J~P,
  Dong X~L, Wang G~W, Bao W~C, Wang Q~H, Yin Z~P, Zhao Z~X and Feng D~L 2018
  {\em Phys. Rev. X\/} {\bf 8} 041056

\bibitem{Sau2010}
Sau J~D, Lutchyn R~M, Tewari S and Das~Sarma S 2010 {\em Phys. Rev. Lett.\/}
  {\bf 104}(4) 040502

\bibitem{Alicea2010}
Alicea J 2010 {\em Phys. Rev. B\/} {\bf 81}(12) 125318

\bibitem{Lutchyn2010}
Lutchyn R~M, Sau J~D and Das~Sarma S 2010 {\em Phys. Rev. Lett.\/} {\bf 105}(7)
  077001

\bibitem{Oreg2010}
Oreg Y, Refael G and von Oppen F 2010 {\em Phys. Rev. Lett.\/} {\bf 105}(17)
  177002

\bibitem{Alicea2011}
Alicea J, Oreg Y, Refael G, von Oppen F and Fisher M~P~A 2011 {\em Nat.
  Phys.\/} {\bf 7} 412--417

\bibitem{Haim2015}
Haim A, Berg E, von Oppen F and Oreg Y 2015 {\em Phys. Rev. Lett.\/} {\bf
  114}(16) 166406

\bibitem{Gorski2018}
G{\'{o}}rski G, Bara{\'{n}}ski J, Weymann I and Doma{\'{n}}ski T 2018 {\em Sci.
  Rep.\/} {\bf 8} 15717

\bibitem{Zhang2019b}
Zhang K, Dong X, Zeng J, Han Y and Qiao Z 2019 {\em Phys. Rev. B\/} {\bf 100}
  045421

\bibitem{Gong2020}
Gong W~J, Gao Z, Li X~S and Zhang L~L 2020 {\em New J. Phys.\/} {\bf 22} 053014

\bibitem{Manousakis2020}
Manousakis J, Wille C, Altland A, Egger R, Flensberg K and Hassler F 2020 {\em
  Phys. Rev. Lett.\/} {\bf 124} 096801

\bibitem{Prada2017}
Prada E, Aguado R and San-Jose P 2017 {\em Phys. Rev. B\/} {\bf 96} 085418

\bibitem{Vijay2016}
Vijay S and Fu L 2016 {\em Phys. Rev. B\/} {\bf 94}(23) 235446

\bibitem{Nayak2008}
Nayak C, Simon S~H, Stern A, Freedman M and Das~Sarma S 2008 {\em Rev. Mod.
  Phys.\/} {\bf 80} 1083--1159

\bibitem{Budich2012}
Budich J~C, Walter S and Trauzettel B 2012 {\em Phys. Rev. B\/} {\bf 85}
  121405(R)

\bibitem{Pedrocchi2015}
Pedrocchi F~L, Bonesteel N~E and DiVincenzo D~P 2015 {\em Phys. Rev. B\/} {\bf
  92}(11) 115441

\bibitem{Mishmash2020}
Mishmash R~V, Bauer B, von Oppen F and Alicea J 2020 {\em Phys Rev B\/} {\bf
  101} 075404

\bibitem{Sekania2017}
Sekania M, Plugge S, Greiter M, Thomale R and Schmitteckert P 2017 {\em Phys.
  Rev. B\/} {\bf 96}(9) 094307

\bibitem{Karzig2013}
Karzig T, Refael G and von Oppen F 2013 {\em Phys. Rev. X\/} {\bf 3} 041017

\bibitem{Scheurer2013}
Scheurer M~S and Shnirman A 2013 {\em Phys. Rev. B\/} {\bf 88} 064515

\bibitem{Posske2020}
Posske T, Chiu C~K and Thorwart M 2020 {\em Phys. Rev. Res.\/} {\bf 2} 023205

\bibitem{Nag2019}
Nag A and Sau J~D 2019 {\em Phys. Rev. B\/} {\bf 100}(1) 014511

\bibitem{Tutschku2020}
Tutschku C, Reinthaler R~W, Lei C, MacDonald A~H and Hankiewicz E~M 2020 {\em
  Phys. Rev. B\/} {\bf 102}(12) 125407

\bibitem{Tutschku2016}
Tutschku C 2016 {\em Topological Quantum Computing Using Nanowire Devices\/}
  Master's thesis Julius-Maximilians-Universit{\"a}t W{\"u}rzburg

\bibitem{Knapp2018}
Knapp C, Karzig T, Lutchyn R~M and Nayak C 2018 {\em Phys. Rev. B\/} {\bf 97}
  125404

\bibitem{Zhang2019c}
Zhang Z~T, Mei F, Meng X~G, Liang B~L and Yang Z~S 2019 {\em Phys. Rev. A\/}
  {\bf 100}(1) 012324

\bibitem{Knapp2016}
Knapp C, Zaletel M, Liu D~E, Cheng M, Bonderson P and Nayak C 2016 {\em Phys.
  Rev. X\/} {\bf 6}(4) 041003

\bibitem{Steiner2020}
Steiner J~F and von Oppen F 2020 {\em Phys. Rev. Res.\/} {\bf 2}(3) 033255

\bibitem{Munk2020}
Munk M~I~K, Schulenborg J, Egger R and Flensberg K 2020 {\em Phys. Rev. Res.\/}
  {\bf 2}(3) 033254

\bibitem{Munk2019}
Munk M~I~K, Egger R and Flensberg K 2019 {\em Phys. Rev. B\/} {\bf 99}(15)
  155419

\bibitem{Qin2019}
Qin L, Li X~Q, Shnirman A and Schön G 2019 {\em New J. Phys.\/} {\bf 21}
  043027

\bibitem{Leijnse2012}
Leijnse M and Flensberg K 2012 {\em Semicond. Sci. Technol.\/} {\bf 27} 124003

\bibitem{Alicea2012}
Alicea J 2012 {\em Rep. Prog. Phys.\/} {\bf 75} 076501

\bibitem{Beenakker2013}
Beenakker C 2013 {\em Annu. Rev. Condens. Matter Phys.\/} {\bf 4} 113--136

\bibitem{Landau2016}
Landau L, Plugge S, Sela E, Altland A, Albrecht S and Egger R 2016 {\em Phys.
  Rev. Lett.\/} {\bf 116} 050501

\bibitem{Plugge2017}
Plugge S, Rasmussen A, Egger R and Flensberg K 2017 {\em New J. Phys.\/} {\bf
  19} 012001

\bibitem{Bonderson2009}
Bonderson P, Freedman M and Nayak C 2009 {\em Ann. Phys.\/} {\bf 324} 787--826

\bibitem{Bonderson2008}
Bonderson P, Freedman M and Nayak C 2008 {\em Phys. Rev. Lett.\/} {\bf 101}
  010501

\bibitem{Chiu2015}
Chiu C~K, Vazifeh M~M and Franz M 2015 {\em EPL\/} {\bf 110} 10001

\bibitem{Wieckowski2020}
Wi{\k{e}}ckowski A, Mierzejewski M and Kupczy{\'{n}}ski M 2020 {\em Phys. Rev.
  B\/} {\bf 101} 014504

\bibitem{Breuer2007}
Breuer H~P and Petruccione F 2007 {\em The Theory of Open Quantum Systems\/}
  (Oxford University Press)

\bibitem{Weiss1999}
Weiss U 1999 {\em Quantum {D}issipative {S}ystems\/} ({World} {Scientific})

\bibitem{May2003}
May V and Kühn O 2003 {\em Charge and Energy Transfer Dynamics in Molecular
  Systems\/} (Wiley-{VCH} Verlag {GmbH})

\bibitem{CohenTannoudji1998}
Cohen-Tannoudji C, Dupont-Roc J and Grynberg G 1998 {\em
  Atom{\textemdash}Photon Interactions\/} (Wiley)

\end{thebibliography}
